# Superconducting joining of melt-textured Y-Ba-Cu-O bulk material

T. Prikhna,[a] W. Gawalek,[b] V. Moshchil,[a] A. Surzhenko,[b] A. Kordyuk,[c]
D. Litzkendorf,[b] S. Dub,[a] V. Melnikov,[d] A. Plyushchay,[c] N. Sergienko,[a]
A. Koval',[e] S. Bokoch,[a] T. Habisreuther [b]

[a] *Institute for Superhard Materials, 2, Avtozavodskaya St, Kiev. 04074, Ukraine*

[b] *Institut fuer Physikalische Hochtechnologie, Winzerlaer Str. 10, D-07743, Jena, Germany*

[c] *Institute of Metal Physics, 36, Vernadsky Ave., 02142, Kiev, Ukraine*

[d] *Institute of Geochemistry, Mineralogy and Ore-Formation, 34, Palladin Ave., Kiev 02142, Ukraine*

[e] *Institute of Problems of Materials Science, 3, Krzhizhanovsky St., Kiev, 02180, Ukraine*

______________________________________________________________________

**Abstract**

The Tm-Ba-Cu-O solder can be successfully used to produce a superconductive joint between MT-YBCO parts. The peculiarities of solidification, phase formation, structure transformations and electromagnetic properties of MT-YBCO soldered with $TmBa_2Cu_3O_{7-\delta}$ are discussed.

______________________________________________________________________

**T.A. Prikhna**, Institute for Superhard Materials of the National Academy of Sciences of Ukraine, 2, Avtozavodskaya Str., 04074 Kiev, Ukraine, Fax.: +380-44-468-86-25,
E-mail: frd@ismanu.kiev.ua

## 1. Introduction

Of known high-temperature superconductors (HTSC) the melt-textured $YBa_2Cu_3O_{7-\delta}$-based (MT-YBCO) is considered as a material to be successfully used for bulk applications like flying wheels, frictionless bearings, electromotors, levitation transport etc. But the devised technologies for production of MT-YBCO don't allow yet large enough high-quality and complex-shaped parts to be produced. So, the problem has arisen of joining MT-YBCO blocks. Is apparent that the superconducting joining is preferable. Kimura et al. [1] proposed to produce SC soldered joints between bulk MT-YBCO blocks using Yb123 SC powder. Walter et al. studied [2] the process of welding MT-YBCO at a pressure of 0.5-MPa using Yb123 and Er123 powders. Zheng et al. [3] have successfully tried Tm123+Y211(25 %) presintered at 1273 K (for ~20 min) as a spacer layer material and obtained a MT-TmBCO seam between MT-YBCO with SC properties comparable to that of MT-YBCO. The main idea is that the incongruent melting temperature of Le123 (Le=Yb, Er, Tm) is lower than that of Y123, so these materials are suitable as solders to joint MT-YBCO parts. The similarity in structures of Le123 (Le=Yb, Er, Tm) and Y123 may result in obtaining of a joint, which is similar in structure and, hence, in SC properties to a bulk material. The structure of MT-YBCO is usually constituted of one or a few (well-oriented relative to each other) large textured SC grains of Y123 with finely dispersed small inclusions of the non-superconductive $Y_2BaCuO_5$ (Y211) phase. The melt–texturing process of Y123 supposes a slow cooling (about 0.3 K/h) in the 1273-1203 K range. To form the soldered seam, the MT-YBCO samples should be repeatedly heated (above 1203 K) that may induce changes of their structure: variation in oxygen content of Y123, decomposition of Y123, recrystallization and coarsening as well as redistribution of Y211, etc. So, it is important to find compromised soldering conditions in order to provide high quality of a joint and not to impair properties of MT-YBCO.

The present paper describes the established peculiarities of a soldered seam formation between MT-YBCO parts, when Tm123 powder is used as a soldering agent, as well as superconductive and mechanical properties of as-soldered MT-YBCO and soldering conditions.

## 2. Experimental

As a starting material for experiments on soldering, we have used top-seeded-melt–textured-grown MT-YBCO cut from 30 mm in diameter blocks (Fig. 1a). Because of the anisotropy of SC behavior it's reasonable to make a joining so that the *ab*-planes of Y123 (along which the critical current is usually stronger than that along the c-axis) of joined parts are parallel. Because of this we have cut MT-YBCO samples perpendicular to the plane, on which the seed was located during melt-texturing (i.e. approximately perpendicular to the *ab*-plane of Y123). The DTA and TG studies show that the temperature of peritectic decomposition of synthesized $TmBa_2Cu_3O_{7-\delta}$ powder is about 1256-1258 K and the decomposition of MT-YBCO occurs at about 1289 K. We have covered MT-YBCO parts to be joint (perpendicular to the *ab*-plane) with a powder either of Tm123 or of a mixture of Tm123 and Y211. The powder was put by sieving or sedimentation from the suspension in acetone. The as-prepared MT-YBCO parts were placed one onto the other and put to the furnace (Fig. 1b). Experiments were done both in air (with subsequent oxygenation in a separate process) and in oxygen.

The sample structure and phase composition have been examined by X-ray, using polarizing optical and scanning electron microscopes. Vickers microhardness was determined employing a Matsuzawa Mod. MXT-70 microhardness tester.

The SC properties, namely the critical current density through the seam $J_{c(s)}$ and through the parts $J_{c(m)}$ of MT-YBCO, were estimated from (1) the field maps (FM) of field-cooled samples using Hall probe and (2) the local levitation forces (LLF). In the case of FM, we used the method based on the Bean model [4]. In the case of LLF, we calculated $J_{c(s)}$ using the equation $J_{c(s)}=[(F_s-F_{cr})/(F_m-F_{cr})]\times J_{c(m)}$, where $J_{c(m)}$ - the critical current density in MT-YBCO (estimated according to the method [5]). Here $F_m$, $F_{cr}$ and $F_s$ - the local levitation forces acting on the spherical magnet (1.5 mm in diameter with 1.9 G×cm$^3$ magnetic moment) at the 0.5 mm-distance above the material before the sample cutting, above the cut, and above the soldered seam, respectively. The sample area that gives ~90% of the measured LLF, is estimated as ~3 mm over the surface and ~1 mm in depth. Thus, LLF method gives us information about the SC properties of the surface, while FM is more suitable to characterize the sample and the seam as a whole. For this reason, when developing the soldering technology, we mainly relied on the FM data.

## 3. Results and discussion

We have succeeded in obtaining SC joints between the MT-YBCO parts with a $J_{c(s)}$ of approximately 1.4 kA/cm$^2$ (FM) or 12 kA/cm$^2$ (LLF).

The results, shown in the Table and Figs. 2, 3 allow us to conclude the following.

Two types of the seam (thick and thin) can be formed (Figs. 2 a,c) with the compatible SC properties (Table, Nos. 5,6). The thick-grained seam (Fig. 1a) usually forms at lower temperatures and when Y211 is added to the Tm123 solder.

The grains of the (Tm,Y)211 phase, the structure of which contains both Tm and Y, have crystallized in the seam during a rather quick cooling (100 K/h) (Figs. 2b, d) and even if Y211 powder have not been added to the solder (Fig. 1b). The amount of 211 phase may increase due to the Y123 partial decomposition at temperatures higher than 1231 K. One can find a phase containing Ba and Cu, but not Y and Tm in the seam as a SEM study shows (Fig. 2d). The addition of Y211 (about 10-25 wt%) has prevented the seam from formation at a soldering temperature above 1263 K and a long holding time (several hours) or low cooling rates (0.5-1 K/h). Even when the sample, in which the Tm123 with 10 wt% of Y211 was used as a solder, has been cooled down in accordance with "melt-textured" regime (regime No. 7 in the Table) the seam cracked after the first cooling in liquid nitrogen. While the sample without Y211 additions (No 7) exhibited good SC properties (Table, Fig. 3).

The heating up to 1273 K (holding time 0.25 h) in air and quick cooling (with furnace: in 1 min samples were cooled from 1257 to 1173 K) leads to a drastic degradation of MT-YBCO SC properties (the level of trapped field decreases by 90 %), while heating up to 1257 K and cooling with the same rate (Table, Nos. 1 and 2) results in less degradation of MT-YBCO trapped field: by 40-50 %. As unit cell parameters of Y123 phase of MT-YBCO were not changed in the experiments Nos. 1,2, we suppose that the MT-YBCO SC properties were decreased due to cracking of the samples as a result of too fast cooling. The soldering in oxygen in accordance with regime Nos. 3-8 (Table) gave us the possibility to retain the SC properties of MT-YBCO. And we haven't observed notable differences between the samples soldered in accordance with regime No. 6 or No. 7 (cooled from 1253 K down to 1228 K at 100 or 0.5 K/h, respectively).

If the rate of heating was too high (400 K/h) the MT-YBCO blocks were warped and cracked.

In spite the fact that the SC properties and mechanical characteristics of the seam are lower than that of the bulk MT-YBCO (see the Table) the Tm123-based SC junction is promising for practical application. SC seam provides better levitation properties of items than non-SC one.

## Conclusions

The best quality junction of MT-YBCO parts was manufactured after heating in $O_2$ up to 1263 K when Tm123 was used as the solder. To find the best soldering conditions, more experiments should be done.

Table

Results of properties study of soldered MT-YBCO samples

| No | Treatment conditions*, solder | $a$, nm $b$, nm $c$, nm (of Y123) | Seam orientation** | Vickers microhardness, GPa (at 1.96-N load) | | Critical current density, kA/cm$^2$ | | |
|---|---|---|---|---|---|---|---|---|
| | | | | | | FM | | LLF |
| | | | | Seam | Material | $J_{c(s)}$ | $J_{c(m)}$ | $J_{c(s)}$ |
| 1 | 1257 K, 0.17 h, in air; fast cooling (with furnace); Tm123+Y211(10 wt%) | 0.3823(3) 0.3891(3) 1.1686(4) | $\perp c$-axis, T | 1.53±0.42 | 3.40±2.48 | 0.4 | 1.3 | 12.0 |
| | | | $\parallel c$-axis | 4.57±0.82 | 3.89±1.90 | 0.7 | 2.0 | 3.9 |
| 2 | Regime ac. to pos. No. 1, Tm123 | 0.3824(2) 0.3884(2) 1.1686(3) | $\perp c$-axis, T | 3.45±0.46 | 4.47±0.64 | 0.4 | 1.6 | 3.2 |
| | | | $\parallel c$-axis | 4.12±0.30 | 4.76±0.72 | 0.9 | 1.6 | 5.7 |
| 3 | 1253 K, 0.5 h, in O$_2$; 100 K/h down to 873 K; Tm123+Y211(10 wt%) | - | $\perp c$-axis, T | 4.09±0.39 | 5.24±0.45 | 1.3 | 3.5 | 1.6 |
| | | | $\parallel c$-axis | 3.43±0.92 | 4.82±0.45 | 1.3 | 3.2 | 7.7 |
| 4 | Regime ac. to pos. No.3 Tm123 | - | $\perp c$-axis, T | 2.26±0.55 | 6.00±0.92 | 0.6 | 1.7 | 8.5 |
| | | | $\parallel c$-axis | 2.93±0.36 | 5.42±0.85 | 1.0 | 3.4 | 6.7 |
| 5 | 1263 K, 0.5 h, in O$_2$; 100 K/h down to 873 K; Tm123+Y211(10 wt%) | - | $\perp c$-axis, T | - | - | 1.0 | 2.0 | 7.2 |
| | | | $\parallel c$-axis | 3.98±0.3 | 4.46±0.9 | 1.4 | 3.7 | 3.9 |
| 6 | Regime ac. to pos. No. 5; Tm123 | 0.3822(2) 0.3890(2) 1.1687(4) | $\perp c$-axis, T | 3.51±0.33 | 5.72±0.93 | 1.2 | 1.9 | 9.4 |
| | | | $\parallel c$-axis | 3.41±0.11 | 5.33±0.56 | 1.4 | 4.6 | 6.1 |
| 7 | 1263 K, 0.25 h, in O$_2$; cooling: 50 K/h to 1253 K, 0.5 K/h to 1228 K, 120 K/h to 873 K; Tm123 (as a sediment) | - | $\perp c$-axis ,T | - | - | 1.0 | 4.2 | - |
| | | | $\perp c$-axis, B | | | 1.4 | 3.4 | - |
| | | | $\parallel c$-axis | - | - | 0.9 | 1.57 | - |
| 8 | Regime ac. to pos. No. 7; Tm123 (sieved) | - | $\perp c$-axis, B | - | - | 0.6 | 1.83 | - |
| | | | $\parallel c$-axis | - | - | 1.0 | 1.8 | - |

\* On cooling to 873 K, samples Nos. 3-8 were cooled down to room temperature by the same multistage procedure.

\** The orientation of the seam is relative: the plane, on which the seed during MT-growth was located, is considered to be parallel ($\parallel$) to the $ab$ plane of Y123 grains, and perpendicular ($\perp$) to the $c$-axis, the «$\parallel$ c» designation means that the seam lies in the plane, which is $\perp$ to the «seed plane», or, approximately, to the $ab$ plane of Y123. T, B - top (seed) and bottom sides of the sample, respectively.

FIGURE CAPTIONS

Fig. 1. The sketches show how the samples have been cut from a MT-YBCO block (a) and the position of a sample in a furnace during soldering (b).

Fig. 2. Soldered seams formed between MT-YBCO parts after heating at 1263 K for 0.5 h in $O_2$, cooling at 100 K/h down to 873 K, multistage cooling for 20 h to room temperature, when ( a,b) Tm123 and (c,d) Tm123+Y211(10 wt%) (see Table, Nos. 6, 5) were used as solders. (a),( c) –microscope pictures in polarized light, (b), (d) – SEM pictures with the relative distributions of Y, Tm, Ba, Cu along the scan line.

Fig. 3. Trapped-field maps for the MT-YBCO sample soldered with Tm123 at 1263 K in $O_2$ (see Table, No. 7).